\documentclass[12pt,manuscript]{aastex}

\shorttitle{Coronal magnetic field}

\shortauthors{Sasikumar et al.}

\begin{document}

\title{An estimate of the magnetic field strength associated with a solar coronal mass ejection from
low frequency radio observations} 

\author{K. Sasikumar Raja\altaffilmark{1}, R. Ramesh\altaffilmark{1}, K. Hariharan\altaffilmark{1}, 
C. Kathiravan\altaffilmark{1}, T. J. Wang\altaffilmark{2}}
\email{sasikumar@iiap.res.in}

\altaffiltext{1}{Indian Institute of Astrophysics, Bangalore 560 034}
\altaffiltext{2}{Department of Physics, The Catholic University of America and 
NASA Goddard Space Flight Center, Code 671, Greenbelt, MD 20771, USA}

\begin{abstract}

We report ground based, low frequency heliograph (80 MHz), spectral (85-35 MHz) and polarimeter (80 and 40 MHz) 
observations of drifting, non-thermal radio continuum  
associated with the `halo' 
coronal mass ejection (CME) that occurred in the solar atmosphere 
on 2013 March 15. 
The magnetic field strengths ($B$) near the radio source
were estimated to be $B \approx 2.2 \pm 0.4$ G at 80 MHz and 
$B \approx 1.4 \pm 0.2$ G at 40 MHz. The corresponding radial distances ($r$) are  
$r \approx 1.9~R_{\odot}$ (80 MHz) and $r \approx 2.2~R_{\odot}$ (40 MHz).

\end{abstract}


\section{Introduction}

It is known that intense, long-lasting (tens of minutes to hours) non-thermal radio continuum  
are observed sometimes in association with the flares and CMEs in the solar atmosphere. 
\citet{Boischot1957} had
designated these events as type IV bursts. Further studies showed that there are two 
classes of type IV bursts:
The first variety occurs after the impulsive phase of the flares 
and drifts in the spectrum to lower frequencies. Interferometer observations indicate that 
the radio source exhibits outward movement through the solar atmosphere with
speeds in the range $\approx$ 200~-~1500 $km~s^{-1}$. Emission can be observed even when the source is located
at large radial distances ($r \approx 5~R_{\odot})$ above the plasma level  
corresponding to the frequency of observation. 
The sources 
have low directivity and are partially circularly polarized. 
The sense of polarization correspond usually to the extraordinary mode (e-mode) 
of the magneto-ionic theory. 
These are called the moving type IV (type IVm) bursts.
The second variety, called the stationary type IV bursts (type IVs), appear near the flare
site during the impulsive phase 
at frequencies $\gtrsim$ 300 MHz and in the post-flare phase
at lower frequencies. 
The type IVs burst is characterized by a source 
whose position does not change and which is located close to or slightly above the plasma level 
corresponding to the frequency of observation. 
The emission 
is circularly polarized usually in the ordinary mode (o-mode) of the  magneto-ionic theory. 
The cone of emission is narrow since type IVs bursts are rarely observed when the associated flare is near 
the limb of the Sun. The high directivity, location close to the plasma level, and the presence of fine structures 
suggest that the emission mechanism is related to the plasma frequency. 
The type IVs bursts at low frequencies may
occur with or without a type IVm burst \citep{Pick1961,Boischot1962,Weiss1963,Wild1963,Stewart1985,Aurass2005,Pick2008}. 

Compared to the type IVs bursts, the type IVm are more closely associated with the CMEs.
It is also possible to infer the CME magnetic field using them. \citet{Gergely1986} had noted that
5\% of all CMEs are associated with type IVm bursts and $33-50$\% of type IVm bursts
are associated with CMEs. But only a few estimates of the field strength have been
reported in the literature. One of the reasons for this is the non-availability of  
simultaneous white light and radio observations close to the Sun ($r \lesssim 2~R_{\odot}$),
and the relatively rare occurrence of the type IVm bursts \cite{White2007}.
The bursts have been explained using either second harmonic 
plasma emission \citep{Duncan1981,Stewart1982b,Gopalswamy1989b,Kundu1989,Ramesh2013} 
or optically thin non-thermal gyro-synchrotron emission  
\citep{Gopalswamy1989a,Bastian1997,Bastian2001,Maia2007,Tun2013}
in the past.
Using simultaneous whitelight, radio heliograph, radio polarimeter and radio spectral observations
we have presented arguments to show that the type IVm burst associated with the
`halo' CME event of 2013 March 15
can be explained on the basis of non-thermal gyro-synchrotron emission
and estimated the magnetic field strength near the source region of the burst.

\section{Observations}

The radio data were obtained 
on 2013 March 15 
at 80 MHz 
with the Gauribidanur RAdioheliograPH \citep{Ramesh1998,Ramesh1999a,Ramesh1999b,Ramesh2006a} in the
imaging mode, the Gauribidanur Radio Interference Polarimeter \citep{Ramesh2008} 
at 80 MHz and 40 MHz in the transit mode,
and over the 85~-~35 MHz band with the Gauribidanur LOw frequency Solar 
Spectrograph \citep{Ebenezer2001,Ebenezer2007,Kishore2014} in the spectral mode. All the 
aforementioned instruments are located in the Gauribidanur radio 
observatory\footnote{http://www.iiap.res.in/centers/radio},
about 100 km north of Bangalore in India \citep{Ramesh2011a}. 
The co-ordinates
of the array are Longitude = $77^{\arcdeg}27^{\arcmin}07^{\arcsec}$ East,
and Latitude = $13^{\arcdeg}36^{\arcmin}12^{\arcsec}$ North.
The GRAPH produces two 
dimensional images of the solar corona with an 
angular resolution of 
$\approx 5^{\arcmin} \times 7^{\arcmin}$ (Right Ascension, R.A. $\times$ Declination, decl.) 
at 80 MHz.
The integration time and the bandwidth of observation are $\approx$ 250 ms
$\approx$ 2 MHz, respectively.
GRIP is an east-west one-dimensional interferometer array and observes the circularly polarized flux 
density from the `whole' Sun at 80 MHz and 40 MHz simultaneously.
Linear polarization, if generated at the correspoding radio source region in the solar atmosphere, 
is presently difficult to detect at low radio frequencies because of the differential Faraday
rotation of the plane of polarization within the typical observing
bandwidths \citep{Grognard1973}.
The GRIP has a broad response pattern (`beam') compared to the Sun in both 
R.A./east-west direction ($\approx 1.5^{\arcdeg}$ at 80 MHz) 
and decl./north-south direction
($\approx 90^{\arcdeg}$). 
So observations with the GRIP in the transit mode essentially reproduces its `east-west beam'
with amplitude proportional to the intensity of the emission from the `whole' Sun, 
weighted by the antenna gain in the corresponding direction.
The integration time and the observing bandwidth are the same as the GRAPH.
The respone pattern of the GLOSS is very broad, 
$\approx 90^{\arcdeg} \times 5^{\arcdeg}$ 
($\rm R.A. \times decl.$) and hence the Sun is a point source for the latter. 
The integration time and the observing bandwidth are comparatively smaller here, $\approx$ 100 ms
and $\approx$ 300 kHz (at each frequency), respectively.
The width of the response pattern of the GLOSS in R.A. (i.e. hour angle) 
is nearly independent of frequency.
The Sun is a point source for both the GRIP and the GLOSS.
The optical data were obtained with the  
Large Angle and 
Spectrometric Coronagraph \citep{Brueckner1995}
onboard the Solar and Heliospheric Observatory (SOHO),
COR1 coronagraph 
of the Sun-Earth Connection Coronal and
Heliospheric Investigation \citep{Howard2008} onboard the Solar 
TErrestrial RElations Observatory (STEREO), and in 193 {\AA} with the 
Atmospheric Imaging Assembly \citep{Lemen2012} onboard the Solar Dynamics
Observatory (SDO).

Figure \ref{fig:figure1} shows the dynamic spectrum obtained
with the GLOSS on 2013 March 15 during the interval 
06:15 - 08:15 UT
in the frequency range $85-35$ MHz.
Two types of enhanced radio emission with differing spectral characteristics
are simultaneously noticeable in the spectrum: 1) a weak stationary continuum
during the period 
$\approx$ 06:30 - 08:10 UT with fine structures, and 2) a comparatively intense 
patch of continuum drifting from 85 MHz to 35 MHz during the period 
$\approx$ 06:55 - 07:50 UT. The fine structures in the background of the latter
are most likely part of the ongoing stationary continuum during the same interval.  
The stationary and the drifting continuum described above are the typical spectral 
signatures of the type IVs and type IVm bursts in the solar atmosphere, 
respectively \citep{Stewart1985}.
The average duration ($\tau$) of the type IVm burst in Figure \ref{fig:figure1} increases with decreasing 
frequency. The increase in the temporal width with decrease in frequency of the region enclosed between 
the `black' lines from 85 MHz to 35 MHz in the spectrum indicates this. The typical widths (i.e. duration) 
are $\tau \approx$ 25 min and $\tau \approx$ 45 min at 85 MHz and 
35 MHz, respectively.
These are consistent with the statistical results on the duration of type IVm bursts
reported by \citet{Robinson1978}. The onset of the burst at 85 MHz is 
$\approx$ 06:55 UT and at 35 MHz is $\approx$ 07:05 UT.

Figure \ref{fig:figure2} shows the time profile 
of the Stokes I \& V radio emission from the solar corona at 80 MHz  
as observed with the GRIP on 2013 March 15.
Similar observations at 40 MHz are shown in Figure \ref{fig:figure3}. 
The observations were carried out in the 
transit mode and hence the observed time profile and their duration essentially corresponds
to the east-west response pattern of the GRIP. 
The peak flux densities, estimated using the polynomial fit to the observations (see the 
overplotted smooth line in Figures \ref{fig:figure2} and \ref{fig:figure3}), 
are $\approx$ 170,000 Jy (Stokes I) and $\approx$ 50,000 Jy (Stokes V) at 80 MHz, and
$\approx$ 181,000 Jy (Stokes I) and $\approx$ 62,000 Jy (Stokes V) at 40 MHz. These values correspond 
mostly to that of the type IVm burst alone since we had subtracted the corresponding mean flux densities 
of the background type IVs burst fine structures using GRIP observations of the same 
outside the type IVm burst period. 
The respective values for the type IVs bursts are $\approx$ 110,000 Jy (Stokes I) 
and $\approx$ 53,000 Jy (Stokes V) at 80 MHz, and
$\approx$ 117,000 Jy (Stokes I) and $\approx$ 61,000 Jy (Stokes V) at 40 MHz. That the flux density of 
particularly the Stokes I emission for the type IVs burst is a significant fraction of that of the 
type IVm burst in the present case is noticeable from their contrast with respect to the background 
in Figure \ref{fig:figure1} also.
The 80 MHz flux densities are reasonably consistent with that reported earlier 
\citep{Kai1969}. 
The spectral index ($\alpha$) between 40 MHz and 80 MHz for the type IVm burst using the above Stokes I
flux densities is $\alpha \approx -0.1$.
Since the non-thermal spectral index is generally $< 0$ \citep{Kraus1986,Subramanian1988}, 
the above value $\alpha \approx -0.1$
indicates that the observed emission in the present case is 
of non-thermal origin.
The estimated degree of
circular polarization ($dcp$) of the type IVm burst in Figures \ref{fig:figure2} and \ref{fig:figure3}
are $\approx 0.29 \pm 0.1$ at 80 MHz and $\approx 0.34 \pm 0.1$ at 40 MHz. Note that due to instrumental limitations,
we observed only $\rm |V|$ with the GRIP and hence $dcp$ = $\rm |V|$/I in the present case. 

The above radio events
were associated with a M1.1 class GOES soft X-ray flare during the interval $\approx$ 05:46~-~08:35 UT
with peak at $\approx$ 06:58 UT, a 1F class H$\alpha$ flare during the interval $\approx$ 06:13~-~08:33 UT with
peak at $\approx$ 06:37 UT from the active region AR 11692 located at N11E12 on the 
solar disk\footnote{swpc.noaa.gov/warehouse/2013.html}, and a 
`halo' CME\footnote{umbra.nascom.nasa.gov/lasco/observations/halo/2013/130315}. 
Figure \ref{fig:figure4} shows the composite of the GRAPH radioheliogram at 80 MHz,
the SOHO-LASCO C2 image and the SDO-AIA 193 {\AA} image, all obtained around $\approx$ 08:00 UT.
Since the observations were during the type IVs burst period in Figure \ref{fig:figure1}, 
the discrete radio source near the disk center is most likely the source region of the type IVs burst. 
Its peak brightness temperature ($T_{b}$) is 
$\approx 3 \times 10^{8}$ K. 
Further details about the type IVs burst shall be reported elsewhere. We will limit 
ourselves to the type IVm burst in the rest of this paper.
We would like to mention here that no GRAPH observations were available during the type IVm burst
in Figure \ref{fig:figure1}. 
The discrete source close to the limb in the north east quadrant in Figure \ref{fig:figure4}
is presumably weak non-thermal radio noise storm activity often observed near the location of
a CME in its aftermath \citep{Kerdraon1983,Kathiravan2007}. The peak $T_{b}$ of the source is $\approx 10^{7}$ K. 

From the movies of the `halo' CME, we find that the its leading edge (LE) was first observed 
in the STEREO-COR1B field of view (FOV) around $\approx$ 06:15 UT  
at $r \approx 1.6~R_{\odot}$. Later, at the onset time of the type IVm burst
at 85 MHz around $\approx$ 06:55 UT (Figure \ref{fig:figure1}), the LE was at 
$r \approx 3.5~R_{\odot}$. This gives a projected linear speed of $\approx 551 ~\rm {km~s^{-1}}$ for the CME LE.
The SOHO-LASCO height-time (h-t) measurements indicate that the CME LE was located 
at $r \approx 4.1~R_{\odot}$ around $\approx$ 07:12 UT\footnote{cdaw.gsfc.nasa.gov}. 
These values are consistent with those extrapolated using the STEREO-COR1B measurements.
We estimated the projected linear speed of the ejecta that moved outwards behind the CME LE 
observed in the STEREO-COR1B FOV (see the asterix marked feature in Figure \ref{fig:figure5}).
The centroid of the ejecta was at $r \approx 1.6~R_{\odot}$ during its first appearance 
at $\approx$ 06:45 UT. Ten minutes later, i.e. at 06:55 UT close to the onset time of the type IVm burst 
at 85 MHz, the ejecta was at $r \approx 1.9~R_{\odot}$
(see Figure \ref{fig:figure5}). 
The above h-t measurements give a speed of $\approx 348 ~\rm {km~s^{-1}}$ for 
the ejecta. This is nearly the same as the speed estimated  
using SOHO-LASCO C2 observations of the same ejecta during $\approx$ 07:24~-~08:24 UT (see Figure \ref{fig:figure4}). 
It appears from Figures \ref{fig:figure4} and \ref{fig:figure5} that the ejecta is close to the plane of the sky 
for both STEREO-COR1B and SOHO-LASCO C2. 
The close agreement between the measured projection speeds of the CME LE and the ejecta
with the above two instruments indicate that the ejecta moved at the same angle 
to the Sun-Earth and Sun-STEREO/B lines. As STEREO-B and Earth were separated by an angle of
about $140^{\circ}$ at the time of the event, we estimate this angle to be $\approx 70^{\circ}$. So
the angle between the ejecta and the plane of the sky for STEREO-COR1B or SOHO-LASCO C2 is  
$\approx 20^{\circ}$. 
 
Any error in the position/size
of the type IVs burst in Figure \ref{fig:figure4} 
due to propagation effects in the solar corona and the Earth's ionosphere
is expected to be minimal ($\approx \pm 0.2~R_{\odot}$) because:
1) positional shifts due to
refraction in the ionosphere is expected to be $\lesssim 0.2~R_{\odot}$ at 80 MHz
in the hour angle range $\pm$~2h \citep{Stewart1982a}. The local
noon at Gauribidanur on 15 March 2013 was around $\approx$ 07:00 UT and
the GRAPH observations described above are within the
above hour angle range; 
2) the effects of scattering 
are considered to be small
at 80 MHz compared to lower frequencies \citep{Aubier1971,Bastian2004,Ramesh2006b}.
High angular resolution 
observations establishing 
that discrete radio sources of angular size $\approx 1^{\arcmin}-3^{\arcmin}$ 
are present in the solar atmosphere from where radio emission at low frequencies 
originate \citep{Kerdraon1979,Lang1987,
Willson1998,Ramesh1999b,Ramesh2000,Ramesh2001,Mercier2006,Kathiravan2011,
Ramesh2012}, ray tracing calculations indicating that the 
turning points of the rays that undergo irregular refraction 
in the solar corona nearly coincide with the location of the plasma (`critical') layer
in the non-scattering case even at high frequencies like 73.8 MHz \citep{Thejappa2008},
that the maximal positional shift (for discrete solar radio sources) due to scattering
is $\lesssim 0.2~R_{\odot}$ at 80 MHz 
\citep{Riddle1974,Robinson1983} also constrain scattering.

\section{Results and Analysis}

\subsection{Emission mechanism}

The type IVm burst in the present case is most likely associated with the CME core like ejecta behind
the CME LE because: 1) the CME LE was located at a large radial distance ($r \approx 3.5~R_{\odot}$) 
at the onset time of the type IVm burst at 85 MHz. The above value is close to the outer limit 
of the radial distance up to which the type IVm bursts have been observed \citep{Smerd1971,Robinson1978};
2) the location of the ejecta ($r \approx 1.9~R_{\odot}$) close to the onset of the type IVm burst at 85 MHz 
around $\approx$ 06:55 UT (see Figure \ref{fig:figure1}) is consistent with the statistical estimate of the radial distance
of the type IVm bursts during their onset at 80 MHz \citep{Smerd1971};  
3) no other bright moving 
structures were noticeable in the STEREO-COR1B FOV during $\approx$ 06:55 - 07:50 UT over $r < 2~R_{\odot}$, 
the interval 
over which the radio emission was observed in Figure \ref{fig:figure1};
4) based on a statistical study of the type IVm bursts at 80 MHz, 
\citet{Gergely1986} had earlier concluded that majority of the type IVm bursts move out with the ejecta behind 
the CME LE. 

Second harmonic plasma emission and/or optically thin gyro-synchrotron emission from mildly 
relativistic electrons have been reported as the likely mechanisms for the type IVm bursts. 
Amongst the above two, we find that the latter is the cause 
in the present case because:
1) \citet{Wild1972,Dulk1973,Melrose1985} had shown that the gyro-synchrotron emission
is strongly suppressed at frequencies $< 2f_{p}$, where $f_{p}$
is the plasma frequency, in the presence of a medium.
We estimated the total coronal electron density (i.e., the density of the background corona and 
the CME together) using STEREO-COR1B pB measurements at the location of the ejecta at $\approx$ 06:55 UT. 
The density of the background corona was determined to beginwith from a pre-CME pB image at $\approx$ 06:05 UT 
using the spherically symmetric 
inversion method \citep{Wang2014}. The electron density of the ejecta was derived from the 
background-subtracted pB radiation averaged over a selected region on the ejecta (see Figure \ref{fig:figure5})
and assuming the line of sight (LOS) column depth of the ejecta equal to its width for pB inversion.
The value is $\rm \approx 7 \times 10^{6}~cm^{-3}$. This corresponds to $f_{p} \approx$ 24 MHz.
Therefore if the type IVm burst of 2013 March 15 were because of  
gyro-synchrotron mechanism, emission at frequencies $\lesssim$ 48 MHz should have been 
progressively weaker at $\approx$ 06:55 UT.
The present observations are consistent with this. The spectral profile of the type IVm 
burst in Figure \ref{fig:figure6}) clearly shows a reduction in the 
observed intensity at frequencies $\lesssim$ 52 MHz.
However at later times the burst is observable at lower frequencies (see Figure \ref{fig:figure1}). 
This is because the ejecta had moved outward in the solar atmosphere
as well as expanded in size with time (see Figures \ref{fig:figure4} and \ref{fig:figure5}).
As a consequence there is a gradual decrease in the total density at the location of the ejecta and 
hence the cut-off frequency for gyro-synchrotron emission. 
Probably this shift in the cut-off towards lower frequencies with time is also responsible 
for the observed drift of the type IVm burst in Figure \ref{fig:figure1}.
Note that the total
density when the ejecta was first observed in the STEREO-COR1B FOV around $\approx$ 06:45 UT 
at $r \approx 1.6~R_{\odot}$ was $\rm \approx 15 \times 10^{6}~cm^{-3}$. Comparing this with the
corresponding measurements at $\approx$ 06:55 UT, we find that the ejecta had moved
a radial distance of $\approx 0.3~R_{\odot}$ in $\approx$ 10 min and the total density
during that period had decreased by about a factor of two. Note that the densities of the 
other rising structures (above the occulter of the coronagraph) of the CME like the `legs' and 
the frontal loop which are comparitively fainter 
(see Figure \ref{fig:figure5}) are $\rm < 7 \times 10^{6}~cm^{-3}$. This indicates
that the corresponding $f_{p} < 24$ MHz. Therefore if the type IVm burst had been due to any
of the aforementioned structures of the CME, the reduction in the intensity of the burst 
at $\approx$ 06:55 UT should have been at lower frequencies than $\approx 52$ MHz. But this is not the case; 
2) the estimated $dcp$ is larger compared to that for reported for type IVm bursts
due to second harmonic plasma emission \citep{Gary1985};
3) the spectral index of the type IVm burst between 40 MHz and 80 MHz as estimated from the GRIP observations
is $\alpha \approx$ -0.1 (see Section 2). 
This is nearly the same as the expected spectral index for gyrosynchrotron emission over the frequency
range $38.5-73.8$ MHz \citep{Gopalswamy1990}.

\subsection{Magnetic field}

\citet{Dulk1985} had shown that for optically thin non-thermal gyro-synchrotron emission, the $dcp$ and $B$ are related
as follows:
\begin{equation}
dcp \approx 1.26 \times 10^{0.035\delta} 10^{-0.071 cos\theta} \left({\frac{f}{f_{B}}}\right)^{-0.782 + 0.545 cos\theta} 
\end{equation}
where $\theta$ is the viewing angle between the LOS
and the magnetic field, $f$ is the frequency of observation, and $f_{B} = 2.8 B$ is
the electron gyro-frequency. The power-law index $\delta$ can be estimated
from the radio flux spectral
index ($\alpha$) through the relationship $\alpha = 1.20-0.90\delta$ \citep{Dulk1985}.
In the present case $\alpha \approx -0.1$. This implies $\delta \approx 1.4$.
The angle between the ejecta and the plane of the sky in the present case is $\approx 20^{\circ}$
(see Section 2). This indicates that the ejecta is nearly normal to the LOS and the 
associated field lines are likely to be radial. 
So we assumed the viewing angle between the LOS
and the magnetic field in the type IVm burst source region
to be the same as the positional angle from the LOS to the ejecta.
Hence $\theta \approx 70^{\circ}$ (see Section 2). 
Substituting for the different parameters in equation (1), we get $B \approx 2.2 \pm 0.4$ G at 80 MHz 
and $B \approx 1.4 \pm 0.2$ G at 40 MHz. The corresponding radial distances are most likely 
$r \approx 1.9~R_{\odot}$ (80 MHz) and $r \approx 2.2~R_{\odot}$ (40 MHz). We estimated this
from the location of the ejecta ($r \approx 1.9~R_{\odot}$) 
during the onset of the type IVm burst at 85 MHz ($\approx$ 06:55 UT), the projected speed of the 
ejecta ($\approx348 ~\rm {km~s^{-1}})$, and the onset of the type IVm burst at 35 MHz ($\approx$ 07:05 UT).
The $dcp$ and the $B$ values are consistent with the results of the model calculations reported 
by \citet{Robinson1974} for $\theta \approx 70^{\circ}$. The sense of polarization is in the e-mode. 
\citet{Gary1985} had remarked that gyro-synchrotron emission is
a possible mechanism at 80 MHz if conditions like $B \approx 2.8$ G at $r \approx 2.5~R_{\odot}$ 
are satisfied. The above estimates of $B$ at 80 MHz in the present case agree reasonably with this.

\section{Summary}

A type IVm radio burst and type IVs radio burst occurred simultaneously on 2013 March 15 
in association with a `halo' CME and a M1.1/1F class  soft X-ray/H$\alpha$ flare.
Radio imaging, spectral and polarimeter observations of the same at low frequencies ($<$ 100 MHz) 
have been reported in this work.
Our results indicate that the type IVm burst can be explained as due to   
optically thin gyro-synchrotron emission
from the non-thermal electrons in the CME core like ejecta behind the CME LE. The estimated magnetic field
strength near the type IVm burst source region is $B \approx 2.2 \pm 0.4$ G and $\approx 1.4 \pm 0.2$ G at 80 MHz
and 40 MHz, respectively. The corresponding radial distances are 
$r \approx 1.9~R_{\odot}$ (80 MHz) and $r \approx 2.2~R_{\odot}$ (40 MHz).
The following results reported earlier indicate that the above values of $B$ 
are plausible:
1) \citet{Dulk1976} estimated the average field strength to be in the range
$\approx 3.1-0.7$ G over $r \approx 1.8-3.1~R_{\odot}$ by assuming gyro-synchrotron mechanism for 
similar CME associated non-thermal radio continuum;
2) \citet{Stewart1982b} reported $B >$ 0.6 G at $r \approx 2.5~R_{\odot}$
for the type IVm burst observed by them at 80 MHz in association with the CME core. The authors
had attributed the radio emission to be either at the fundamental or the second
harmonic of the plasma frequency. Note that if the density requirements for particularly
the second harmonic plasma emission are nearly the same as that for the gyro-synchrotron emission,
then it is possible that the corresponding $B$ values could be similar \citep{Dulk1976};
3) \citet{Gopalswamy1989a,Bastian2001} 
evaluated $B \approx 1.5$ G at
$r \approx 1.5~R_{\odot}$ based on similar non-thermal radio continuum due to gyro-synchrotron 
emission from the associated CMEs;
4) {low frequency (77 MHz) polarimeter observations of a coronal streamer associated radio source 
indicate that magnetic field in the former 
at $r \approx 1.7~R_{\odot}$ is $\approx$ 5 G.
The present estimates of $B$ are reasonably consistent with this. 
That both the CMEs and streamers are primarily
density enhancements in the solar atmosphere could be a reason for this;
5) for a similar type IVm burst event explained on the basis of
optically thin gyro-synchrotron emission from 
the mildly relativistic non-thermal electrons in the magnetic 
field of the associated CME core, \citet{Tun2013} showed that 
$B \approx 5-15$ G at $r \approx 1.7~R_{\odot}$;
6) type IVm radio bursts associated with the `leg' of the corresponding CMEs
and generated due to second
harmonic plasma emission from the enhanced electron density there
indicate that $B \approx$ 4 G at $r \approx 1.6~R_{\odot}$ \citep{Ramesh2013}.
Considering that the coronal magnetic field associated 
with the active regions have a range of values \citep{Dulk1978,Ramesh2003,
Ramesh2011b,Sasikumar2013},  
the different estimates mentioned above can
be regarded as reasonable. 
With measurements of the coronal magnetic field being very limited, particularly in 
close association with a CME, the results indicate that contemporaneous whitelight and
radio observations of the solar corona close to the Sun ($r \lesssim 2~R_{\odot}$) 
are desirable to understand the CMEs and the associated magnetic field.

\acknowledgements

It is a pleasure to thank the staff of the Gauribidanur observatory
for their help in observations, maintenance of the antenna 
and receiver systems there. 
We also thank the referee for his/her comments that helped to bring out the
results more clearly.
The SOHO data are
produced by a consortium of the Naval Research Laboratory (USA),
Max-Planck-Institut fuer Aeronomie (Germany),~Laboratoire
d'Astronomie (France),~and the University of Birmingham (UK). SOHO
is a project of international cooperation between ESA and NASA.
The SOHO-LASCO CME catalog and STEREO movies are generated and maintained at the CDAW Data Center
by NASA and the Catholic University of America in cooperation with the
Naval Research Laboratory.
The SDO/AIA data are courtsey of the NASA/SDO and the AIA
science teams. The work of TJW was supported by 
NASA Cooperative Agreement NNG11PL10A to CUA and NASA grant NNX12AB34G.

\begin{figure}
\epsscale{0.80}
\plotone{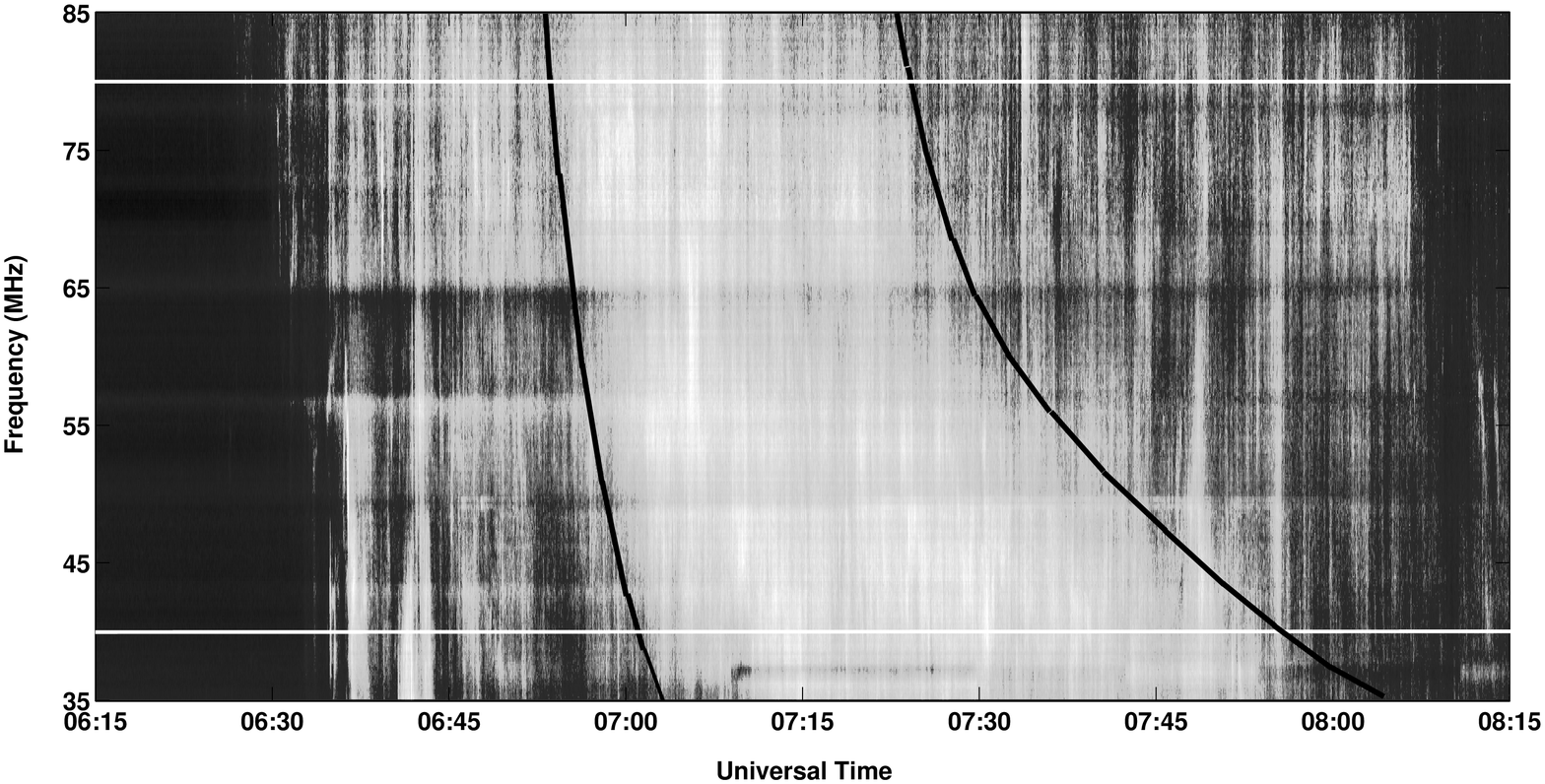}
\caption{Dynamic spectrum of the solar radio emission observed with the GLOSS
on 2013 March 15 during 
06:15 - 08:15 UT
in the frequency range $85-35$ MHz. The stationary
emission during the interval 
$\approx$ 06:30 - 08:10 UT and the drifting emission during $\approx$ 06:55 - 07:50 UT
correspond respectively to the type IVs and type IVm bursts mentioned in the text.
The two `white' horizontal lines indicate 40 MHz and 80 MHz portion of the spectrum. 
The other horizontal line like features noticeable in the spectrum, for eg. near $\approx$ 55 MHz, 
$\approx$ 65 MHz, etc. are due to local radio frequency interference (RFI).
The two
slanted `black' lines indicate the approximate interval over which the type IVm burst was observed at different 
frequencies.}
\label{fig:figure1}
\end{figure}

\clearpage

\begin{figure}
\epsscale{0.80}
\plotone{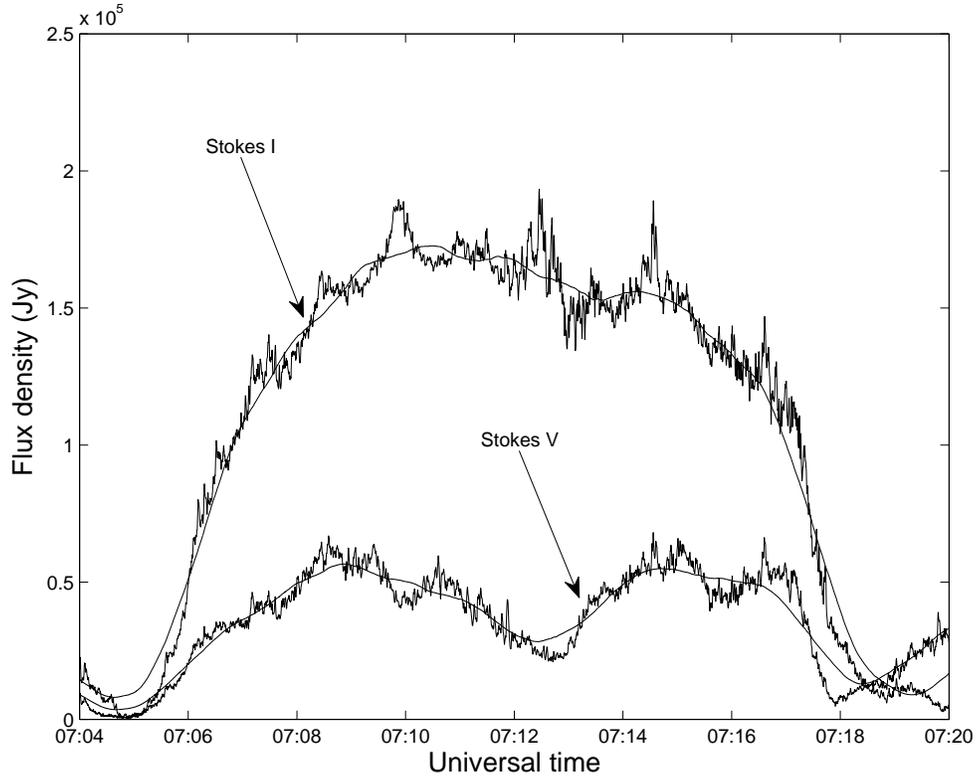}
\caption{Time profile of the Stokes I and Stokes V emission observed with the GRIP 
at 80 MHz on 2013 March 15 in the transit mode. The duration of the observations
correspond approximately to the width of the response pattern of 
the GRIP at 80 MHz in the east-west direction (see Section 2).
The overplotted smooth line is the polynomial fit to the observations.}
\label{fig:figure2}
\end{figure}

\clearpage

\begin{figure}
\epsscale{0.80}
\plotone{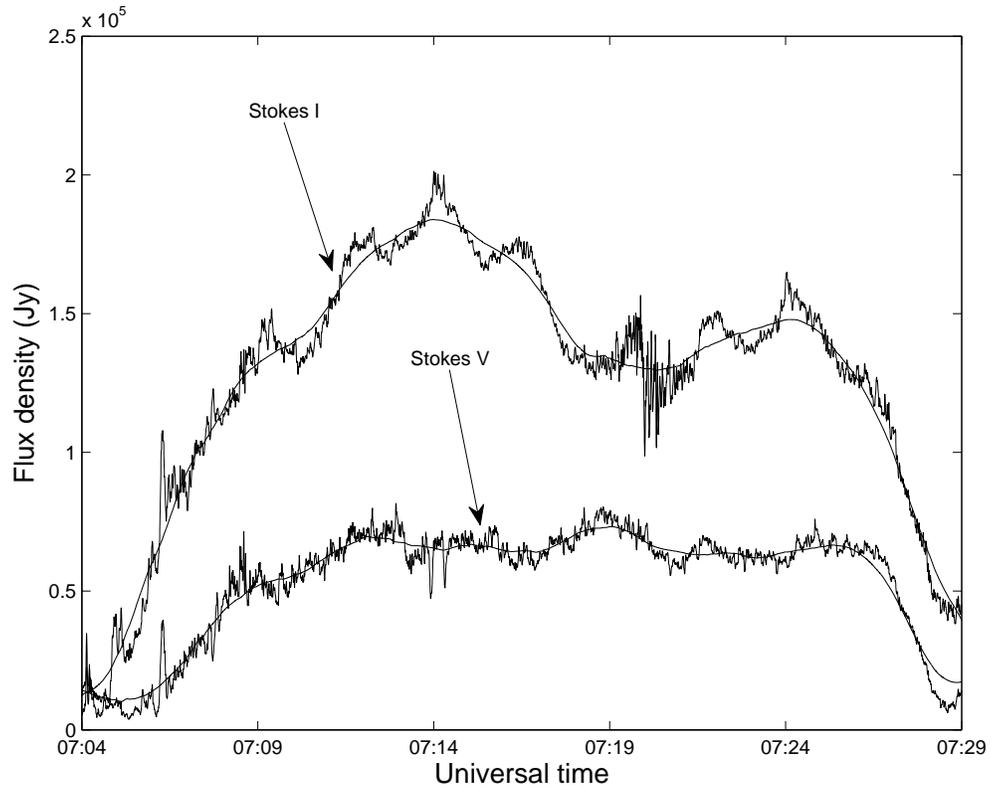}
\caption{Same as Figure \ref{fig:figure2} but at 40 MHz.}
\label{fig:figure3}
\end{figure}

\clearpage

\begin{figure}
\epsscale{0.80}
\plotone{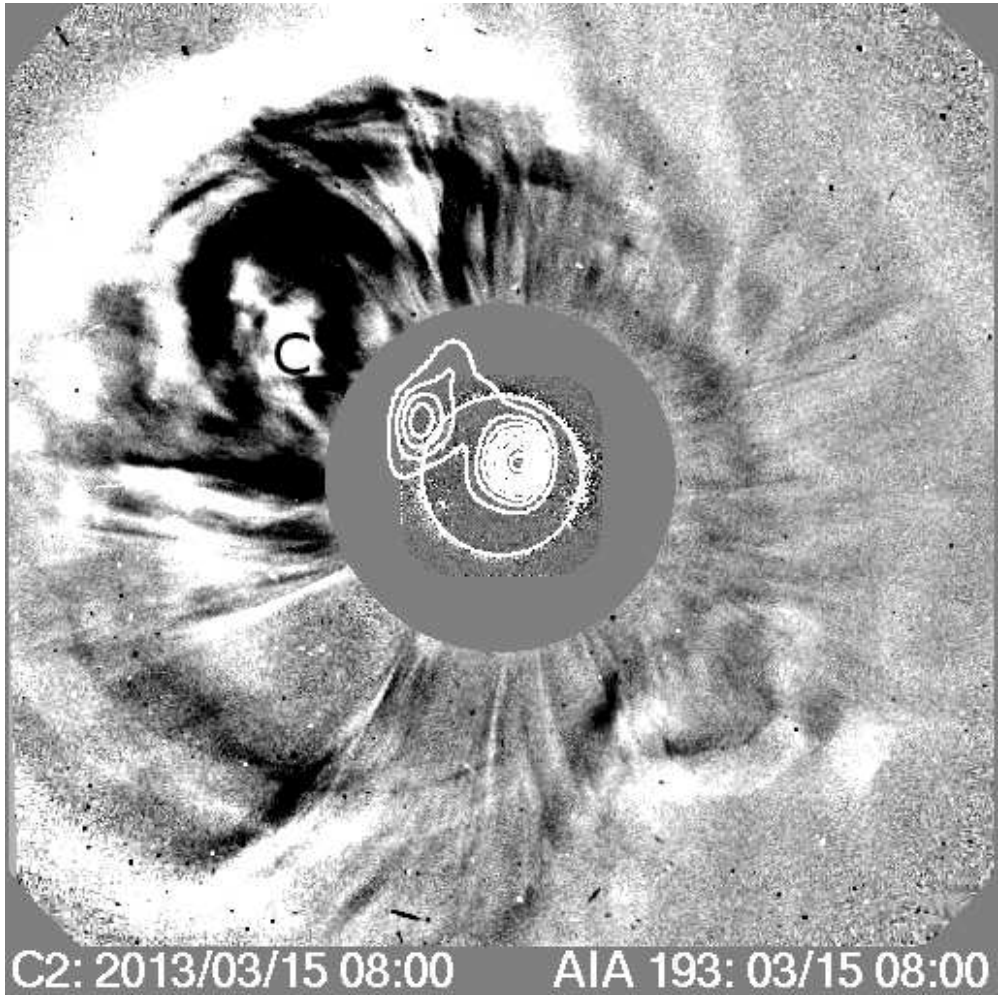}
\caption{A composite of the 80 MHz GRAPH radioheliogram of the 
type IVs burst observed on 2013 March 15 around $\approx$ 08:00 UT 
(contours in white colour) and the
SOHO-LASCO C2, SDO-AIA (193 {\AA}) images obtained 
close to the same time that day. The discrete source
of radio emission near the disk center is the type IVs burst mentioned
in the text. The `white' circle(radius = $1~R_{\odot}$) 
at the center indicates the solar limb. The bigger, concentric `grey' circle (radius $\approx 2.2~R_{\odot}$) 
represents the occulting disk of the SOHO-LASCO C2 coronagraph. Solar north is straight up and solar east 
is to the left in the image. The white light feature marked `C' is the CME core like ejecta mentioned in the
text.} 
\label{fig:figure4}
\end{figure}

\clearpage

\begin{figure}
\epsscale{0.80}
\plotone{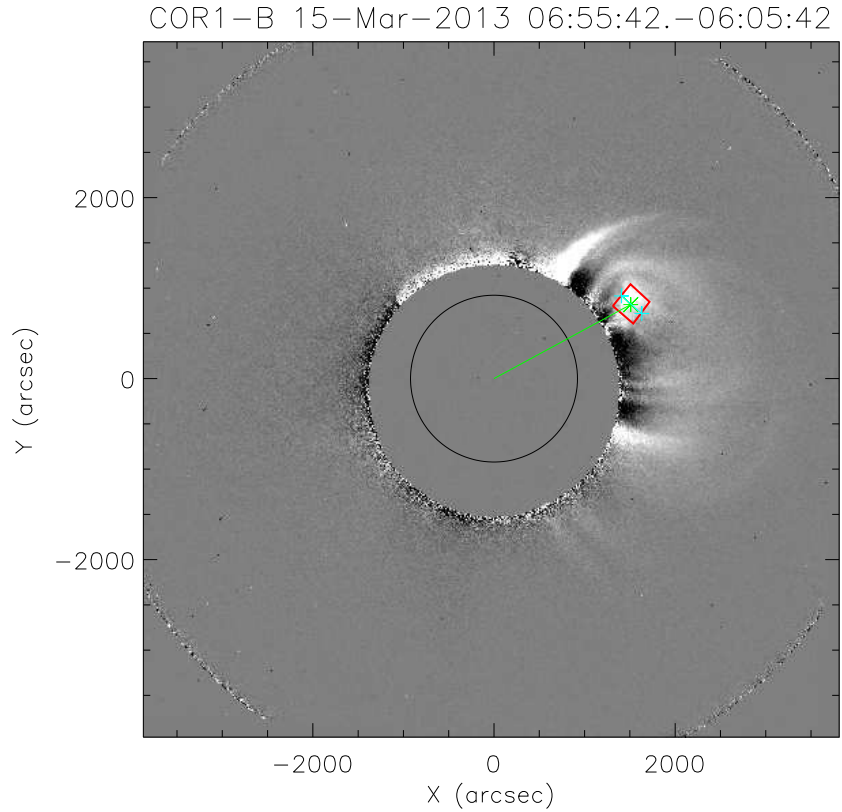}
\caption{STEREO-COR1B pB difference image obtained on 2013 March 15 around 06:55 UT.
The subtracted reference image was observed at 06:05 UT prior to the CME onset.
The region marked with a rectangular box is used for measuring the density of the CME ejecta. 
The 'grey' circle (radius  $\approx 1.4~R_{\odot}$) represents the occulting disk of the coronagraph. 
The asterisk within the rectangular box marks the same 
feature (ejecta) marked 'C' in Figure \ref{fig:figure4}.}
\label{fig:figure5}
\end{figure}

\clearpage

\begin{figure}
\epsscale{0.80}
\plotone{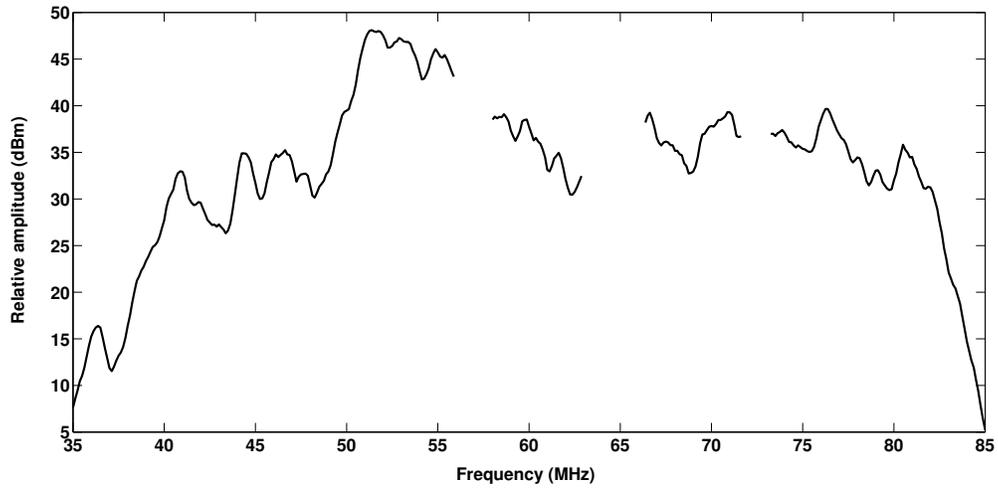}
\caption{Spectral profile of the type IVm burst in Figure \ref{fig:figure1} at $\approx$ 06:55 UT.
The gaps in the profile near $\approx$ 55 MHz, $\approx$ 65 MHz, etc. correspond to the frequency channels 
affected by RFI.}
\label{fig:figure6}
\end{figure}

\end{document}